\documentclass[DIV=calc, paper=a4, fontsize=11pt, twocolumn]{article}	 

\usepackage[total={7in, 8.5in}]{geometry}
\usepackage[english]{babel} 
\usepackage[protrusion=true,expansion=true]{microtype} 

\usepackage[sc]{mathpazo} 
\usepackage[T1]{fontenc} 
\usepackage{amsmath,amsfonts,amsthm} 
\usepackage{gensymb}
\usepackage[svgnames]{xcolor} 
\usepackage[hang, small,labelfont=bf,up,textfont=it,up]{caption} 
\usepackage{booktabs} 
\usepackage{fix-cm}	 

\usepackage{sectsty} 
\allsectionsfont{\usefont{OT1}{phv}{b}{n}} 

\usepackage[absolute]{textpos}

\usepackage{tikz}
\usepackage{pgfplots} 
\usetikzlibrary{patterns}

\usepackage{graphicx}
\graphicspath{{figures/}}
\usepackage{hyperref}
\usepackage{subcaption}

\usepackage{fancyhdr} 
\usepackage{lastpage} 
\usepackage[none]{hyphenat}
\usetikzlibrary{arrows.meta}
\usepackage{comment}

\newcommand{\JFadd}[1]{\textcolor{black}{#1}}

\newcommand{\rev}[1]{\textcolor{black}{{#1}}}

\usepackage{titling} 

\pretitle{\vspace{-30pt} \begin{center} \fontsize{30}{30} \usefont{OT1}{phv}{b}{n} \color{Black} \selectfont} 

	\title{Fluctuations and arctic curve in the Aztec diamond
	} 
	
	\posttitle{\par\end{center}\vskip 0.5em} 

\preauthor{\begin{center}\large \lineskip 0.5em \usefont{OT1}{phv}{b}{sl} \color{Black}} 
	
	\author{Bryan Debin$^{1}$, Jean-François de Kemmeter$^{2}$, Philippe Ruelle$^{1}$} 
	
	\postauthor{\vspace{10pt}\footnotesize \usefont{OT1}{phv}{m}{sl} \color{Black} 
		\\$^{2}$ Department of Mathematics and Namur Institute for Complex Systems (naXys),
		University of Namur, Rue de Bruxelles 61, Namur, B-5000, Belgium\\
		jean-francois.dekemmeter@unamur.be
		\\ $^{1}$ Institut de Recherche en Mathématique et Physique,
		Université catholique de Louvain, Louvain-la-Neuve, B-1348, Belgium  
		\date{\today} 
		
		\par\end{center}} 

\begin{document}
	
	\twocolumn[
	\begin{@twocolumnfalse}
		\maketitle
		\vspace{-20pt}
		
		\begin{abstract}
			\textbf{Domino tilings of Aztec diamonds are known to exhibit an arctic phenomenon, namely a separation between frozen regions (in which all the dominoes have the same orientation) and a central disordered region (where dominoes are found without any apparent order). This separation was proved to converge, under a suitable rescaling, to the Airy process whose $1$-point distribution is the Tracy-Widom distribution. In this work, we conjecture, by means of numerical analysis, that the boundary between the frozen and disordered regions, converges, for the same rescaling, to the Airy line ensemble, a generalisation of the Airy process.}
		\end{abstract}
	\end{@twocolumnfalse}
	\vspace{20pt}
	]
	
	

\section*{Introduction}
\rev{
A variety of patterns can be observed in Nature. They often result from the interactions among microscopic units. An example is the adsorption of some molecules on graphite \cite{blunt2008random}. When adsorbed, these molecules bind to each other such that the angle formed between any two neighboring molecules is either $60 \degree$ or $120 \degree$. As revealed by scanneling tunneling microscope, the resulting entropically stabilized configurations are equivalent to rhombus \textit{tilings}. Tilings and more specifically domino tilings will be at the heart of this article. Tiling models are also closely related to \textit{vertex} models, the latter being first introduced by Linus Pauling to explain the residual entropy of ice at zero temperature \cite{pauling1935structure}. In those models, boundary conditions can strongly influence the bulk properties of the system, a feature recently highlighted experimentally in a colloïdal artificial ice \cite{rodriguez2021topological}. In the following, we introduce the notion of domino tiling and discuss the impact of boundary conditions through the celebrated \textit{Aztec diamond}. 
}

\section*{Domino tilings of a rectangle}

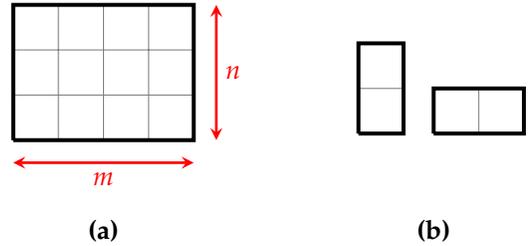
\begin{figure}
	\centering
		\centering
		\begin{tikzpicture}[scale=0.6]
		\draw[step=1cm,color=gray] (0,0) grid (4,3);
		\draw[color=black,line width=1.5pt] (0,0) -- (4,0) -- (4,3)--(0,3)--(0,0);
		\draw[stealth-stealth, line width = 1, color=red] (0,-0.5) -- (4,-0.5);
		\draw (2,-0.5) node[below,red] {$m$};
		\draw[stealth-stealth, line width = 1, color=red] (4.5,0) -- (4.5,3);
		\draw (4.5,1.5) node[right,red] {$n$};
		\draw (2,-2) node {\textbf{(a)}};
		\end{tikzpicture}\hspace{1cm}
		\begin{tikzpicture}[scale=0.6]
		\begin{minipage}{0.2\textwidth}
		\draw[step=1cm,color=gray] (0,0) grid (1,2);
		\draw[color=black,line width=1.5pt] (0,0) -- (1,0) -- (1,2)--(0,2)--(0,0);
		\end{minipage}\hspace{1cm}
		\begin{minipage}{0.2\textwidth}
		\draw[step=1cm,color=gray] (0,0) grid (2,1);
		\draw[color=black,line width=1.5pt] (0,0) -- (2,0) -- (2,1)--(0,1)--(0,0);
		\end{minipage}
		\draw (0,-2) node {\textbf{(b)}};
		\end{tikzpicture}
	\caption{(a) A rectangulair domain of size $4\times 3$ tileable by dominoes. (b) A domino, be its orientation vertical or horizontal, is the union of two unit squares. }
	\label{fig:rectangle_dominoes}
\end{figure}

Imagine a tiler who wishes to tile a rectangular domain of dimension $m\times n$ ($m,n \in \mathbb{N}$). He has at its disposal rectangular tiles of dimension $1\times 2$, here and now named dominoes, see Figure \ref{fig:rectangle_dominoes}.

\begin{figure}
	\centering
		\begin{tikzpicture}[scale=0.4]
		\draw[step=1cm,color=gray] (0,0) grid (4,2);
		\draw[color=black,line width=1.5pt] (0,0) -- (4,0) -- (4,2)--(0,2)--(0,0);
		\draw (2,0) node[below] {$Z_{2,n}$};
		\draw (6,1) node {$\equiv$};
		\end{tikzpicture}\hspace{0.5cm}
		\begin{tikzpicture}[scale=0.4]
		\draw[step=1cm,color=gray] (0,0) grid (4,2);
		\draw[color=black,line width=1.5pt] (0,0) -- (4,0) -- (4,2)--(0,2)--(0,0);
		\draw[color=black,line width=1.5pt] (0,0) -- (1,0) -- (1,2)--(0,2)--(0,0);
		\draw[fill=black!30!white] (0,0) rectangle (1,2);
		\draw[stealth-stealth, line width = 1, color=red] (1,2.5) -- (4,2.5);
		\draw (2.5,2.5) node[above, color=red] {$n-1$};
		\draw (2,0) node[below] {$Z_{2,n-1}$};
		\draw (6,1) node {$+$};
		\end{tikzpicture}\hspace{0.5cm}
		\begin{tikzpicture}[scale=0.4]
		\draw[step=1cm,color=gray] (0,0) grid (4,2);
		\draw[color=black,line width=1.5pt] (0,0) -- (4,0) -- (4,2)--(0,2)--(0,0);
		\draw[color=black,line width=1.5pt] (0,1) -- (2,1) ;
		\draw[color=black,line width=1.5pt] (2,0) -- (2,2) ;
		\draw[fill=black!30!white] (0,0) rectangle (2,1);
		\draw[fill=black!30!white] (0,1) rectangle (2,2);
		\draw[stealth-stealth, line width = 1, color=red] (2,2.5) -- (4,2.5);
		\draw (3,2.5) node[above, color=red] {$n-2$};
		\draw (2,0) node[below] {$Z_{2,n-2}$};
		\end{tikzpicture}
	\caption{The number $Z_{2,n}$ of domino tilings of a $2\times n$ rectangle can be decomposed into two sums, depending on the orientation of the domino that covers the leftmost unit squares. This leads to the Fibonacci recurrence relation $Z_{2,n} = Z_{2,n-1} + Z_{2,n-2}$.}
	\label{fig:Fibonnaci}
\end{figure}
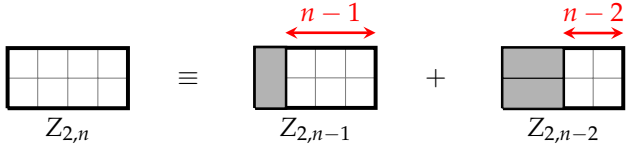

Before he gets down to work, the tiler would like to answer the following question: is the domain tileable and, if so, how many distinct tilings are there ? 
The domain is said \textit{tileable} if there exists at least one tiling, namely a configuration for which any point of the domain is covered by \textit{exactly} one domino, such that no domino crosses the boundary of the domain. Let $Z_{m,n}$ be the number of tilings of such domain. A necessary (and actually sufficient) condition for this domain to be tileable, i.e. $Z_{m,n}\geq 1$, is that $m\cdot n$ must be even (and non-zero). As an example, let us consider the case $m=2$, for which an explicit formula can be readily obtained. The first few terms of the sequence ${\big(Z_{2,n}\big)}_n$  are $0,1,2,3,5,8,\cdots$.
The reader might have recognised the Fibonacci sequence. Indeed, as depicted in Figure \ref{fig:Fibonnaci}, the terms $Z_{2,n}$ satisfy the following recurrence relation:
\begin{equation} 
\begin{split}
&Z_{2,n} = Z_{2,n-1} + Z_{2,n-2}, \\
&Z_{2,0} = 0 \,,\, Z_{2,1} = 1.
\end{split}
\end{equation}

\begin{figure}
	\centering
	\includegraphics[width=\columnwidth]{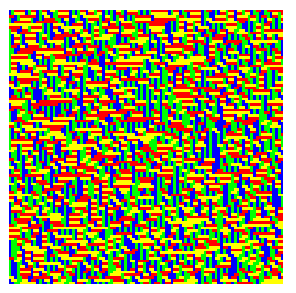}
	\caption{Configuration of a square of size $n=m=100$, randomly sampled, using the Janvresse algorithm \cite{janvresse2006shuffling}.}
	\label{fig:config_rectangle}
\end{figure}

Solving this recurrence relation leads to the explicit formula for the number of domino tilings of a $2\times n$ rectangle:
\begin{equation}
Z_{2,n} = \frac{1}{\sqrt{5}} { \left(\frac{1+\sqrt{5}}{2} \right)}^n -\frac{1}{\sqrt{5}} { \left(\frac{1-\sqrt{5}}{2}\right) }^n.
\end{equation}
An explicit formula also exists for general values of $m$ and $n$, athough its derivation is much more involved than the case $m=2$. It was proved that \cite{kasteleyn1961statistics,temperley1961dimer}:
\begin{equation}
Z_{m,n} = \prod_{j=1}^{\lceil{m/2}\rceil} \prod_{k=1}^{\lceil{n/2}\rceil}
\left(
4\cos^2 \frac{\pi j}{m+1} + 4 \cos^2 \frac{\pi k}{n+1}
\right).
\end{equation}
The number of tilings grows quite rapidly with $m$ and $n$, provided $m,n$ are not both odd integers. For example, if $m=n$, the above formula gives $Z_{2,2} = 2, Z_{4,4}=36, Z_{6,6}=6728, Z_{8,8}=12988816$. Figure \ref{fig:config_rectangle} shows a configuration, for $m=n=100$, randomly chosen among all the tilings. \JFadd{Let us for now make abstraction of the colours attributed to the dominoes.}

\begin{figure}
	\centering
	\begin{tikzpicture}[scale=0.4]
	\foreach \i in {0,1,2,3}
	{
		\draw[step=1cm,color=gray] (-\i,\i) grid (\i+2,\i+1);
		\draw[step=1cm,color=gray] (-3+\i,3+\i) grid (3-\i+2,3+\i+1);
	}
	\draw[color=black,line width=1.5pt] (0,0)--(2,0)--(2,1)--(3,1)--(3,2)--(4,2)--(4,3)--(5,3)--(5,4)--(4,4)--(4,5)--(3,5)--
	(3,6)--(2,6)--(2,7)--(0,7)--(0,6)--(-1,6)--(-1,5)--(-2,5)--(-2,4)--(-3,4)--(-3,3)--
	(-2,3)--(-2,2)--(-1,2)--(-1,1)--(0,1)--(0,0);
	\draw[stealth-stealth, line width = 1, color=red] (1,3) -- (1,7);
	\node[color=red] (n) at (1.5, 4.5) {$n$};
	\node[] (n) at (1, -1) {\textbf{(a)}};
	\end{tikzpicture}\hspace{1cm}
	\begin{tikzpicture}[scale=0.4]
	\foreach \i in {0,1,2,3}
	{
		\draw[step=1cm,color=gray] (-\i,\i) grid (\i+2,\i+1);
		\draw[step=1cm,color=gray] (-3+\i,4+\i) grid (3-\i+2,4+\i+1);
	}
	\draw[step=1cm, draw=gray, fill=black!20!white] (-3,3) grid (3+2,3+1);
	\draw[color=black,line width=1.5pt] (0,0)--(2,0)--(2,1)--(3,1)--(3,2)--(4,2)--(4,3)--(5,3)--(5,5)--(4,5)--(4,6)--(3,6)--
	(3,7)--(2,7)--(2,8)--(0,8)--(0,7)--(-1,7)--(-1,6)--(-2,6)--(-2,5)--(-3,5)--(-3,3)--
	(-2,3)--(-2,2)--(-1,2)--(-1,1)--(0,1)--(0,0);
	\draw[stealth-stealth, line width = 1, color=red] (1,4.1) -- (1,8);
	\node[color=red] (n) at (1.5, 5.5) {$n$};
	\draw[color=blue] (1,4)  node{$\bullet$} ;
	\draw[color=blue] (1.5,3.5) node {$O$};
	\node[] (n) at (1, -1) {\textbf{(b)}};
	\end{tikzpicture}
	\caption{(a) The unique tiling of the domain consists exclusively of horizontal dominoes. By adding a row of $2n$ unit squares to this domain, we obtain the Aztec diamond (\textbf{AD}) of order $n$ shown in right (b). The \textbf{AD} is the set of unit squares whose centres $(i,j)$ satisfy the inequality $\vert i \vert + \vert j \vert \leq n$ with the origin $O$ taken to be the centre of the \textbf{AD}. }
	\label{fig:AD}
\end{figure}
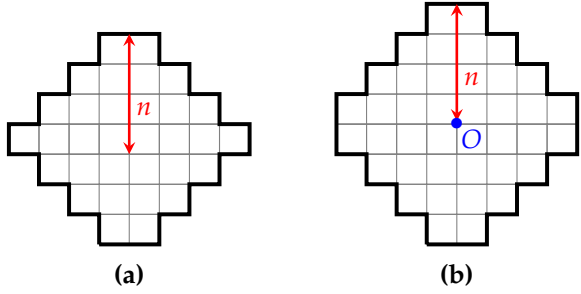

As can be seen, any macroscopic portion of this domain contains, on average, the same fraction of vertical dominoes than horizontal ones. One might wonder whether similar observations hold true for any other tileable domain. The answer to this question is no. Take for instance the domain given in Figure \ref{fig:AD} (a), whose dimension is parametrised by $n$. 

The domain is tileable but the constraints induced by the boundary are so strong that there is only one configuration, the one for which all the dominoes are placed horizontally. 

\section*{Aztec diamond and arctic phenomenon}
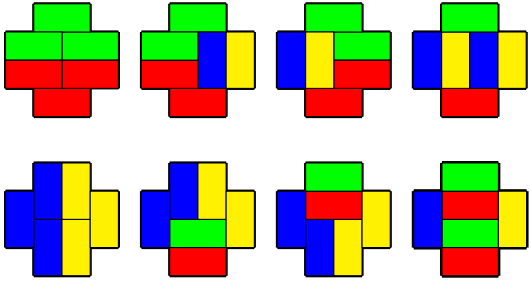
\begin{figure}
	\begin{center}
		\begin{minipage}{0.15\linewidth}
			\begin{center}
				\begin{tikzpicture}[scale=1.5]
				\draw[very thick,color=black] (0.25,0.75) -- (0.75,0.75);
				\draw[very thick,color=black] (0.25,-0.25) -- (0.75,-0.25);
				\draw[very thick,color=black] (0,0) -- (0,0.5);
				\draw[very thick,color=black] (1,0) -- (1,0.5);
				\draw[very thick,color=black] (0,0.5) -- (0.25,0.5);
				\draw[very thick,color=black] (0.25,0.5) -- (0.25,0.75);
				\draw[very thick,color=black] (0.75,0.75) -- (0.75,0.5);
				\draw[very thick,color=black] (0.75,0.5) -- (1,0.5);
				\draw[very thick,color=black] (0,0) -- (0.25,0);
				\draw[very thick,color=black] (0.25,0) -- (0.25,-0.25);
				\draw[very thick,color=black] (0.75,-0.25) -- (0.75,0);
				\draw[thick,color=black] (0.75,0) -- (1,0);
				\draw[thick,color=black] (0.25,0.5) -- (0.75,0.5);
				\draw[thick,color=black] (0,0.25) -- (1,0.25);
				\draw[thick,color=black] (0.25,0) -- (0.75,0);
				\draw[thick,color=black] (0.5,0) -- (0.5,0.5);	
				\draw[fill = red]  (0.25,-0.25) rectangle (0.75,0);
				\draw[fill = green]  (0.5,0.25) rectangle (1,0.5);
				\draw[fill = red]  (0.5,0) rectangle (1,0.25);
				\draw[fill = green]  (0.25,0.5) rectangle (0.75,0.75);
				\draw[fill = green]  (0,0.25) rectangle (0.5,0.5);
				\draw[fill = red]  (0,0) rectangle (0.5,0.25);
				\end{tikzpicture}
			\end{center}
		\end{minipage}\hspace{0.5cm}%
		\begin{minipage}{0.15\linewidth}
			\begin{center}
				\begin{tikzpicture}[scale=1.5]
				\draw[very thick,color=black] (0.25,0.75) -- (0.75,0.75);
				\draw[very thick,color=black] (0.25,-0.25) -- (0.75,-0.25);
				\draw[very thick,color=black] (0,0) -- (0,0.5);
				\draw[very thick,color=black] (1,0) -- (1,0.5);
				\draw[very thick,color=black] (0,0.5) -- (0.25,0.5);
				\draw[very thick,color=black] (0.25,0.5) -- (0.25,0.75);
				\draw[very thick,color=black] (0.75,0.75) -- (0.75,0.5);
				\draw[very thick,color=black] (0.75,0.5) -- (1,0.5);
				\draw[very thick,color=black] (0,0) -- (0.25,0);
				\draw[very thick,color=black] (0.25,0) -- (0.25,-0.25);
				\draw[very thick,color=black] (0.75,-0.25) -- (0.75,0);
				\draw[very thick,color=black] (0.75,0) -- (1,0);
				\draw[thick,color=black] (0.25,0.5) -- (0.75,0.5);
				\draw[thick,color=black] (0,0.25) -- (0.5,0.25);
				\draw[thick,color=black] (0.75,0) -- (0.75,0.5);
				\draw[thick,color=black] (0.25,0) -- (0.75,0);
				\draw[thick,color=black] (0.5,0) -- (0.5,0.5);
				\draw[fill = red]  (0.25,-0.25) rectangle (0.75,0);
				\draw[fill = green]  (0.25,0.5) rectangle (0.75,0.75);
				\draw[fill = blue]  (0.5,0) rectangle (0.75,0.5);
				\draw[fill = yellow]  (0.75,0) rectangle (1,0.5);
				\draw[fill = red]  (0,0) rectangle (0.5,0.25);
				\draw[fill = green]  (0,0.25) rectangle (0.5,0.5);
				\end{tikzpicture}
			\end{center}
		\end{minipage}\hspace{0.5cm}%
		\begin{minipage}{0.15\linewidth}
			\begin{center}
				\begin{tikzpicture}[scale=1.5]
				\draw[very thick,color=black] (0.25,0.75) -- (0.75,0.75);
				\draw[very thick,color=black] (0.25,-0.25) -- (0.75,-0.25);
				\draw[very thick,color=black] (0,0) -- (0,0.5);
				\draw[very thick,color=black] (1,0) -- (1,0.5);
				\draw[very thick,color=black] (0,0.5) -- (0.25,0.5);
				\draw[very thick,color=black] (0.25,0.5) -- (0.25,0.75);
				\draw[very thick,color=black] (0.75,0.75) -- (0.75,0.5);
				\draw[very thick,color=black] (0.75,0.5) -- (1,0.5);
				\draw[very thick,color=black] (0,0) -- (0.25,0);
				\draw[very thick,color=black] (0.25,0) -- (0.25,-0.25);
				\draw[very thick,color=black] (0.75,-0.25) -- (0.75,0);
				\draw[very thick,color=black] (0.75,0) -- (1,0);
				\draw[thick,color=black] (0.25,0.5) -- (0.75,0.5);
				\draw[thick,color=black] (0.5,0.25) -- (1,0.25);
				\draw[thick,color=black] (0.25,0) -- (0.25,0.5);
				\draw[thick,color=black] (0.25,0) -- (0.75,0);
				\draw[thick,color=black] (0.5,0) -- (0.5,0.5);
				\draw[fill = red]  (0.25,-0.25) rectangle (0.75,0);
				\draw[fill = green]  (0.5,0.25) rectangle (1,0.5);
				\draw[fill = red]  (0.5,0) rectangle (1,0.25);
				\draw[fill = green]  (0.25,0.5) rectangle (0.75,0.75);
				\draw[fill = blue]  (0,0) rectangle (0.25,0.5);
				\draw[fill = yellow]  (0.25,0) rectangle (0.5,0.5);
				\end{tikzpicture}
			\end{center}
		\end{minipage}\hspace{0.5cm}%
		\begin{minipage}{0.15\linewidth}
			\begin{center}
				\begin{tikzpicture}[scale=1.5]
				\draw[very thick,color=black] (0.25,0.75) -- (0.75,0.75);
				\draw[very thick,color=black] (0.25,-0.25) -- (0.75,-0.25);
				\draw[very thick,color=black] (0,0) -- (0,0.5);
				\draw[very thick,color=black] (1,0) -- (1,0.5);
				\draw[very thick,color=black] (0,0.5) -- (0.25,0.5);
				\draw[very thick,color=black] (0.25,0.5) -- (0.25,0.75);
				\draw[very thick,color=black] (0.75,0.75) -- (0.75,0.5);
				\draw[very thick,color=black] (0.75,0.5) -- (1,0.5);
				\draw[very thick,color=black] (0,0) -- (0.25,0);
				\draw[very thick,color=black] (0.25,0) -- (0.25,-0.25);
				\draw[very thick,color=black] (0.75,-0.25) -- (0.75,0);
				\draw[very thick,color=black] (0.75,0) -- (1,0);
				\draw[thick,color=black] (0.25,0.5) -- (0.75,0.5);
				\draw[thick,color=black] (0.25,0) -- (0.25,0.5);
				\draw[thick,color=black] (0.75,0) -- (0.75,0.5);
				\draw[thick,color=black] (0.25,0) -- (0.75,0);
				\draw[thick,color=black] (0.5,0) -- (0.5,0.5);
				\draw[fill = red]  (0.25,-0.25) rectangle (0.75,0);
				\draw[fill = yellow]  (0.25,0) rectangle (0.5,0.5);
				\draw[fill = blue]  (0.5,0) rectangle (0.75,0.5);
				\draw[fill = green]  (0.25,0.5) rectangle (0.75,0.75);
				\draw[fill = blue]  (0,0) rectangle (0.25,0.5);
				\draw[fill = yellow]  (0.75,0) rectangle (1,0.5);
				\end{tikzpicture}
			\end{center}
		\end{minipage}\vspace{0.5cm}

		\begin{minipage}{0.15\linewidth}
			\begin{center}
				\begin{tikzpicture}[scale=1.5]
				\draw[very thick,color=black] (0.25,0.75) -- (0.75,0.75);
				\draw[very thick,color=black] (0.25,-0.25) -- (0.75,-0.25);
				\draw[very thick,color=black] (0,0) -- (0,0.5);
				\draw[very thick,color=black] (1,0) -- (1,0.5);
				\draw[very thick,color=black] (0,0.5) -- (0.25,0.5);
				\draw[very thick,color=black] (0.25,0.5) -- (0.25,0.75);
				\draw[very thick,color=black] (0.75,0.75) -- (0.75,0.5);
				\draw[very thick,color=black] (0.75,0.5) -- (1,0.5);
				\draw[very thick,color=black] (0,0) -- (0.25,0);
				\draw[very thick,color=black] (0.25,0) -- (0.25,-0.25);
				\draw[very thick,color=black] (0.75,-0.25) -- (0.75,0);
				\draw[very thick,color=black] (0.75,0) -- (1,0);
				\draw[thick,color=black] (0.25,0.25) -- (0.75,0.25);
				\draw[thick,color=black] (0.25,0) -- (0.25,0.5);
				\draw[thick,color=black] (0.75,0) -- (0.75,0.5);
				\draw[thick,color=black] (0.5,0.25) -- (0.5,0.75);
				\draw[thick,color=black] (0.5,-0.25) -- (0.5,0.25);
				\draw[thick,color=black] (0.25,0) -- (0.75,0);
				\draw[fill = blue]  (0.25,-0.25) rectangle (0.5,0.25);
				\draw[fill = yellow]  (0.5,-0.25) rectangle (0.75,0.25);
				\draw[fill = blue]  (0.25,0.25) rectangle (0.5,0.75);
				\draw[fill = yellow]  (0.5,0.25) rectangle (0.75,0.75);
				\draw[fill = blue]  (0,0) rectangle (0.25,0.5);
				\draw[fill = yellow]  (0.75,0) rectangle (1,0.5);
				\end{tikzpicture}
			\end{center}
		\end{minipage}\hspace{0.5cm}%
		\begin{minipage}{0.15\linewidth}
			\begin{center}
				\begin{tikzpicture}[scale=1.5]
				\draw[very thick,color=black] (0.25,0.75) -- (0.75,0.75);
				\draw[very thick,color=black] (0.25,-0.25) -- (0.75,-0.25);
				\draw[very thick,color=black] (0,0) -- (0,0.5);
				\draw[very thick,color=black] (1,0) -- (1,0.5);
				\draw[very thick,color=black] (0,0.5) -- (0.25,0.5);
				\draw[very thick,color=black] (0.25,0.5) -- (0.25,0.75);
				\draw[very thick,color=black] (0.75,0.75) -- (0.75,0.5);
				\draw[very thick,color=black] (0.75,0.5) -- (1,0.5);
				\draw[very thick,color=black] (0,0) -- (0.25,0);
				\draw[very thick,color=black] (0.25,0) -- (0.25,-0.25);
				\draw[very thick,color=black] (0.75,-0.25) -- (0.75,0);
				\draw[very thick,color=black] (0.75,0) -- (1,0);
				\draw[thick,color=black] (0.25,0.25) -- (0.75,0.25);
				\draw[thick,color=black] (0.25,0) -- (0.25,0.5);
				\draw[thick,color=black] (0.75,0) -- (0.75,0.5);
				\draw[thick,color=black] (0.5,0.25) -- (0.5,0.75);
				\draw[thick,color=black] (0.25,0) -- (0.75,0);
				\draw[fill = red]  (0.25,-0.25) rectangle (0.75,0);
				\draw[fill = green]  (0.25,0) rectangle (0.75,0.25);
				\draw[fill = blue]  (0.25,0.25) rectangle (0.5,0.75);
				\draw[fill = yellow]  (0.5,0.25) rectangle (0.75,0.75);
				\draw[fill = blue]  (0,0) rectangle (0.25,0.5);
				\draw[fill = yellow]  (0.75,0) rectangle (1,0.5);
				\end{tikzpicture}
			\end{center}
		\end{minipage}\hspace{0.5cm}%
		\begin{minipage}{0.15\linewidth}
			\begin{center}
				\begin{tikzpicture}[scale=1.5]
				\draw[very thick,color=black] (0.25,0.75) -- (0.75,0.75);
				\draw[very thick,color=black] (0.25,-0.25) -- (0.75,-0.25);
				\draw[very thick,color=black] (0,0) -- (0,0.5);
				\draw[very thick,color=black] (1,0) -- (1,0.5);
				\draw[very thick,color=black] (0,0.5) -- (0.25,0.5);
				\draw[very thick,color=black] (0.25,0.5) -- (0.25,0.75);
				\draw[very thick,color=black] (0.75,0.75) -- (0.75,0.5);
				\draw[very thick,color=black] (0.75,0.5) -- (1,0.5);
				\draw[very thick,color=black] (0,0) -- (0.25,0);
				\draw[very thick,color=black] (0.25,0) -- (0.25,-0.25);
				\draw[very thick,color=black] (0.75,-0.25) -- (0.75,0);
				\draw[very thick,color=black] (0.75,0) -- (1,0);
				\draw[thick,color=black] (0.25,0.25) -- (0.75,0.25);
				\draw[thick,color=black] (0.25,0) -- (0.25,0.5);
				\draw[thick,color=black] (0.75,0) -- (0.75,0.5);
				\draw[thick,color=black] (0.5,-0.25) -- (0.5,0.25);
				\draw[thick,color=black] (0.25,0.5) -- (0.75,0.5);
				\draw[fill = blue]  (0.25,-0.25) rectangle (0.5,0.25);
				\draw[fill = yellow]  (0.5,-0.25) rectangle (0.75,0.25);
				\draw[fill = red]  (0.25,0.25) rectangle (0.75,0.5);
				\draw[fill = green]  (0.25,0.5) rectangle (0.75,0.75);
				\draw[fill = blue]  (0,0) rectangle (0.25,0.5);
				\draw[fill = yellow]  (0.75,0) rectangle (1,0.5);
				\end{tikzpicture}
			\end{center}
		\end{minipage}\hspace{0.5cm}%
		\begin{minipage}{0.15\linewidth}
			\begin{center}
				\begin{tikzpicture}[scale=1.5]
				\draw[line width=1.5pt,color=black] (0.25,0.75) -- (0.75,0.75);
				\draw[line width=1.5pt,color=black] (0.25,-0.25) -- (0.75,-0.25);
				\draw[line width=1.5pt,color=black] (0,0) -- (0,0.5);
				\draw[line width=1.5pt,color=black] (1,0) -- (1,0.5);
				\draw[line width=1.5pt,color=black] (0,0.5) -- (0.25,0.5);
				\draw[line width=1.5pt,color=black] (0.25,0.5) -- (0.25,0.75);
				\draw[line width=1.5pt,color=black] (0.75,0.75) -- (0.75,0.5);
				\draw[line width=1.5pt,color=black] (0.75,0.5) -- (1,0.5);
				\draw[line width=1.5pt,color=black] (0,0) -- (0.25,0);
				\draw[line width=1.5pt,color=black] (0.25,0) -- (0.25,-0.25);
				\draw[line width=1.5pt,color=black] (0.75,-0.25) -- (0.75,0);
				\draw[line width=1.5pt,color=black] (0.75,0) -- (1,0);
				\draw[very thick,color=black] (0.25,0.25) -- (0.75,0.25);
				\draw[very thick,color=black] (0.25,0) -- (0.25,0.5);
				\draw[very thick,color=black] (0.75,0) -- (0.75,0.5);
				\draw[very thick,color=black] (0.25,0) -- (0.75,0);
				\draw[very thick,color=black] (0.25,0.5) -- (0.75,0.5);
				\draw[fill = red]  (0.25,-0.25) rectangle (0.75,0);
				\draw[fill = green]  (0.25,0) rectangle (0.75,0.25);
				\draw[fill = red]  (0.25,0.25) rectangle (0.75,0.5);
				\draw[fill = green]  (0.25,0.5) rectangle (0.75,0.75);
				\draw[fill = blue]  (0,0) rectangle (0.25,0.5);
				\draw[fill = yellow]  (0.75,0) rectangle (1,0.5);
				\end{tikzpicture}
			\end{center}
		\end{minipage}
	\end{center}
	\caption{The \textbf{AD} of order $2$ can be tiled in $8$ distinct ways.}
	\label{8configurations_n2}
\end{figure}

The example shown in Figure \ref{fig:AD} (a) is in some sense pathological and not so interesting per se. A much more interesting situation is obtained by slightly modifying the domain, through the introduction of a row of unit squares, as shown in Figure \ref{fig:AD} (b). This slight modification has drastic consequences as we shall see in the following. This domain is known as the \textit{Aztec diamond} (\textbf{AD} in the following) and was first introduced in \cite{elkies1992ASM_AD1,elkies1992ASM_AD2}. Formally, it is the set of unit squares whose centres $(i,j)$ are such that $\vert i \vert + \vert j \vert \leq n$, with the origin $(0,0)$ coinciding with the centre of the \textbf{AD}. Notice already that the domain is symmetric under a quarter-turn rotation.  For $n=2$, there are eight configurations shown in Figure \ref{8configurations_n2}. 

\begin{figure*}
	\centering
	\begin{subfigure}[b]{0.28\textwidth}
		\includegraphics[width=\columnwidth]{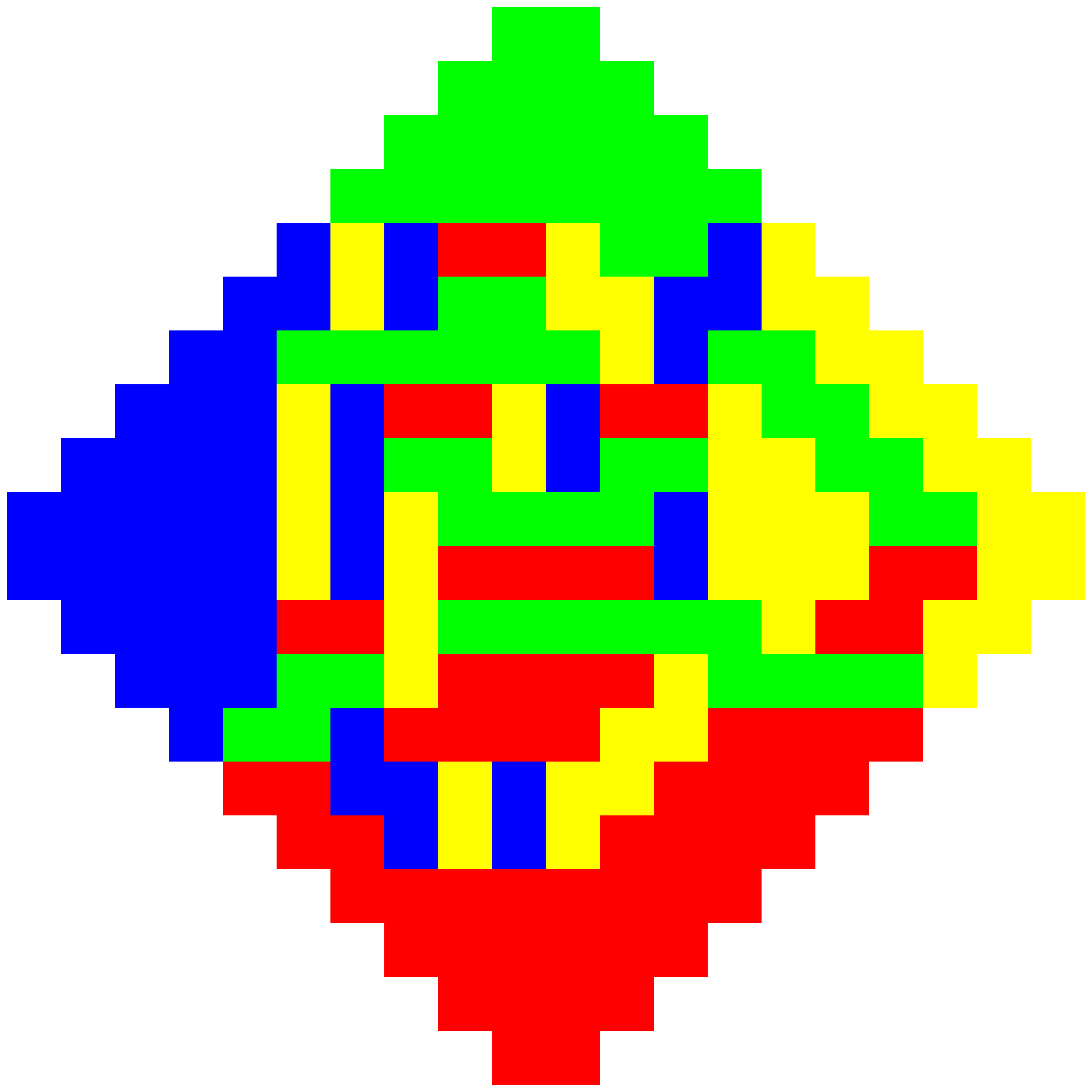}
	\end{subfigure}\hspace{1cm}
	\begin{subfigure}[b]{0.28\textwidth}
		\includegraphics[width=\columnwidth]{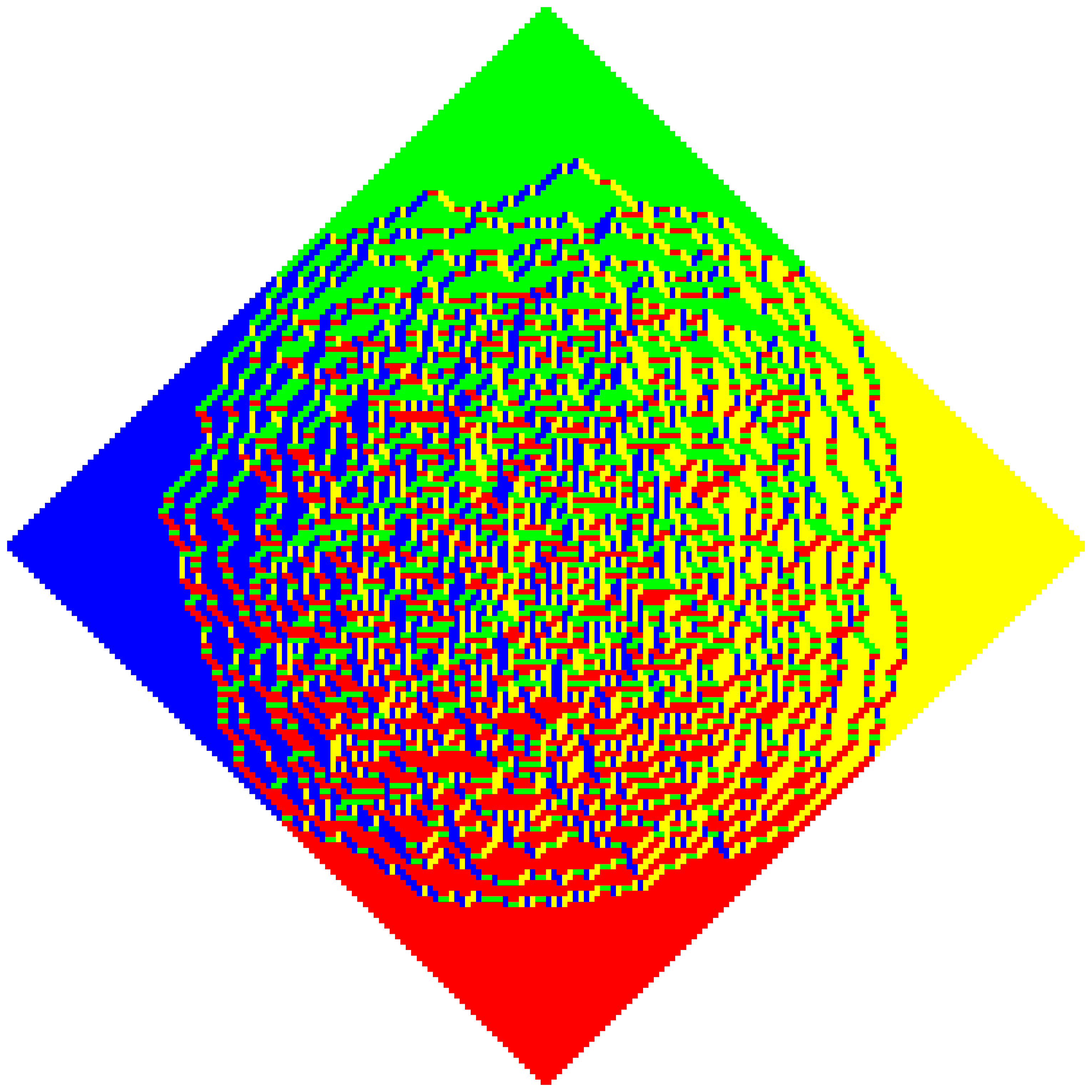}
	\end{subfigure}\hspace{1cm}
	\begin{subfigure}[b]{0.28\textwidth}
		\includegraphics[width=\columnwidth]{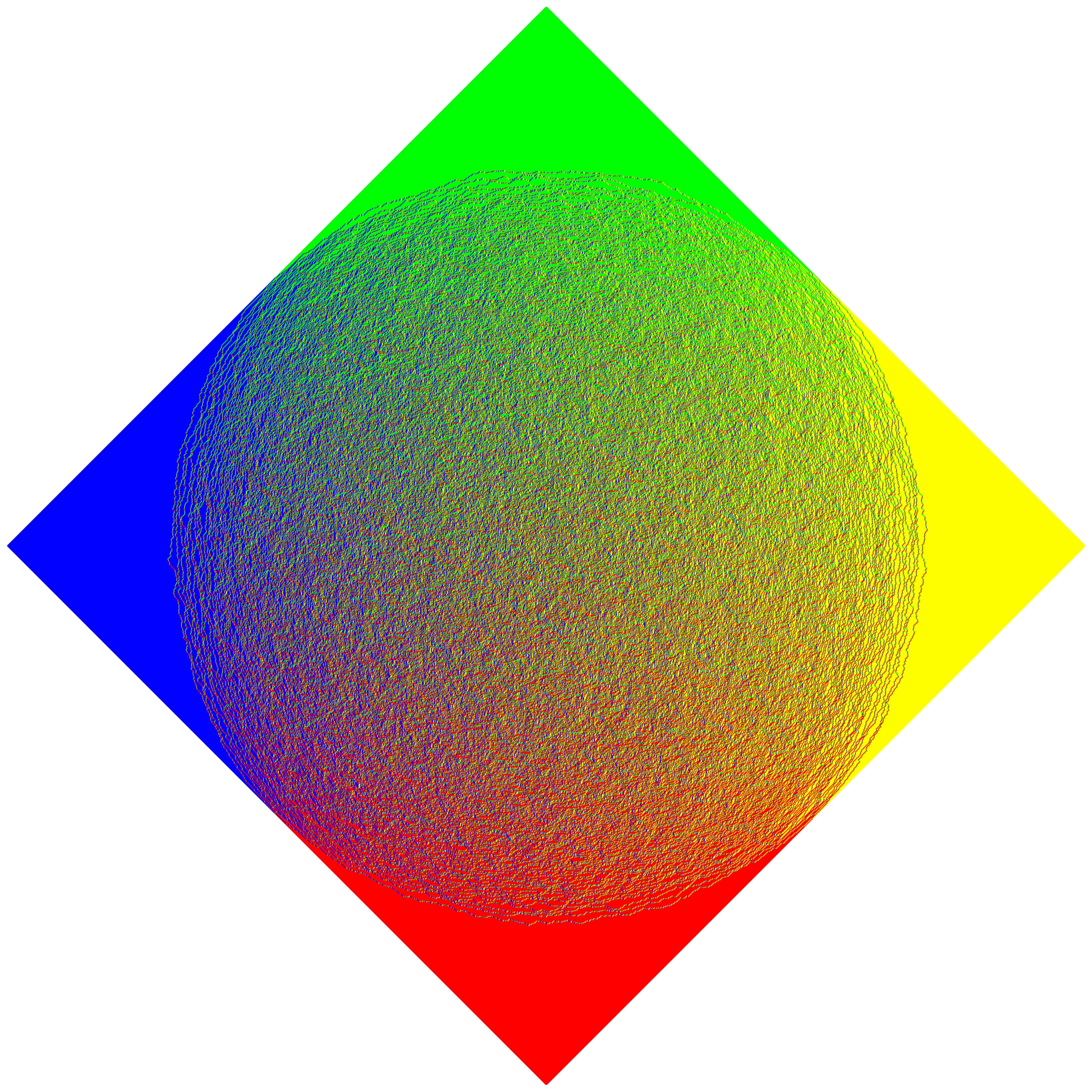}
	\end{subfigure}
	\caption{Configurations of an \textbf{AD} of order $10$, $100$ and $1000$ (from left to right), generated by the shuffling algorithm.}
	\label{fig:config_AD}
\end{figure*}

Let $A_n$ ($n \in \mathbb{N}_0$) be the number of distinct configurations of an \textbf{AD} of order $n$. It was proved that:
\begin{equation}
A_n = 2^{n(n+1)/2}
\end{equation}
Although remarkable by its simplicity, none of the many known proofs is elementary. Maybe even more surprising, tilings of an \textbf{AD} exhibit a peculiar behaviour as $n$ increases, as can be seen in Figure \ref{fig:config_AD} which shows configurations sampled at random for orders $n=10, n=100$ and $n=1000$.

We observe that the four corners of the \textbf{AD} look like brickwalls: each of them contains exclusively horizontal or vertical dominoes. In contrast, far away from the boundaries, horizontal and vertical dominoes alternate in a disordered fashion. As the order of the \textbf{AD} increases, the separation between the four frozen corners and the central disordered region becomes sharper. In the limit $n\rightarrow + \infty$, this separation is a circle, known as the arctic curve of the model. It means that, with probability $1$, each of the four regions outside the circle is covered by the same type of dominoes. This may be heuristically understood as follows. 

Suppose that the unit square centred at $(-n+1/2,-1/2)$ is covered by an horizontal domino, see Figure \ref{fig:constraintsAD} (a). This forces all the dominoes adjacent to the northwest and southwest boundaries to be horizontal too.%
\begin{figure}
	\centering
	\begin{tikzpicture}[scale=0.4]
	\foreach \i in {0,1,2,3}
	{
		\draw[step=1cm,color=gray] (-\i,\i) grid (\i+2,\i+1);
		\draw[step=1cm,color=gray] (-3+\i,4+\i) grid (3-\i+2,4+\i+1);
	}
	\draw[step=1cm, draw=gray, fill=black!20!white] (-3,3) grid (3+2,3+1);
	\draw[step=1cm,color=gray, fill = black!30!white] (-3,3) grid (-1,4) rectangle (-3,3);
	\draw[step=1cm,color=gray, fill = red!30!white] (-3,4) grid (-1,5) rectangle (-3,4);
	\draw[step=1cm,color=gray, fill = red!30!white] (-2,5) grid (0,6) rectangle (-2,5);
	\draw[step=1cm,color=gray, fill = red!30!white] (-1,6) grid (1,7) rectangle (-1,6);
	\draw[step=1cm,color=gray, fill = red!30!white] (0,7) grid (2,8) rectangle (0,7);
	
	\draw[step=1cm,color=gray, fill = red!30!white] (-2,3) grid (0,2) rectangle (-2,3);
	\draw[step=1cm,color=gray, fill = red!30!white] (-1,2) grid (1,1) rectangle (-1,2);
	\draw[step=1cm,color=gray, fill = red!30!white] (0,1) grid (2,0) rectangle (0,1);
	\draw[color=black,line width=1.5pt] (0,0)--(2,0)--(2,1)--(3,1)--(3,2)--(4,2)--(4,3)--(5,3)--(5,5)--(4,5)--(4,6)--(3,6)--
	(3,7)--(2,7)--(2,8)--(0,8)--(0,7)--(-1,7)--(-1,6)--(-2,6)--(-2,5)--(-3,5)--(-3,3)--
	(-2,3)--(-2,2)--(-1,2)--(-1,1)--(0,1)--(0,0);
	\draw[black, very thick] (-1,2)--(0,2)--(0,3)--(-1,3)--(-1,5)--(0,5)--(0,6);
	\draw[black, very thick] (-3,3)--(-1,3);
	\draw[black, very thick] (-3,4)--(-1,4);
	\draw[black, very thick] (-2,5)--(0,5);
	\draw[black, very thick] (0,6)--(1,6)--(1,7)--(0,7);
	\draw[black, very thick] (1,7)--(2,7);
	\draw[black, very thick] (0,2)--(1,2)--(1,1);
	\draw[black, very thick] (0,1)--(2,1);
	\draw[black, very thick] (-1,6)--(0,6);
	\draw (1,-1) node {\textbf{(a)}}; 
	\end{tikzpicture}\hspace{1cm}
	\begin{tikzpicture}[scale=0.4]
	\foreach \i in {0,1,2,3}
	{
		\draw[step=1cm,color=gray] (-\i,\i) grid (\i+2,\i+1);
		\draw[step=1cm,color=gray] (-3+\i,4+\i) grid (3-\i+2,4+\i+1);
	}
	\draw[step=1cm, draw=gray, fill=black!20!white] (-3,3) grid (3+2,3+1);
	\draw[step=1cm,color=gray, fill = black!30!white] (-3,3) grid (-2,5) rectangle (-3,3);
	\draw[step=1cm,color=gray, fill = black!30!white] (-2,4) grid (0,5) rectangle (-2,4);
	\draw[step=1cm,color=gray, fill = black!30!white] (-2,3) grid (0,4) rectangle (-2,3);
	\draw[step=1cm,color=gray, fill = black!30!white] (-2,5) grid (0,6) rectangle (-2,5);
	\draw[step=1cm,color=gray, fill = red!30!white] (-1,6) grid (1,7) rectangle (-1,6);
	\draw[step=1cm,color=gray, fill = red!30!white] (0,7) grid (2,8) rectangle (0,7);
	
	\draw[step=1cm,color=gray, fill = black!30!white] (-2,3) grid (0,2) rectangle (-2,3);
	\draw[step=1cm,color=gray, fill = red!30!white] (-1,2) grid (1,1) rectangle (-1,2);
	\draw[step=1cm,color=gray, fill = red!30!white] (0,1) grid (2,0) rectangle (0,1);
	\draw[color=black,line width=1.5pt] (0,0)--(2,0)--(2,1)--(3,1)--(3,2)--(4,2)--(4,3)--(5,3)--(5,5)--(4,5)--(4,6)--(3,6)--
	(3,7)--(2,7)--(2,8)--(0,8)--(0,7)--(-1,7)--(-1,6)--(-2,6)--(-2,5)--(-3,5)--(-3,3)--
	(-2,3)--(-2,2)--(-1,2)--(-1,1)--(0,1)--(0,0);
	\draw[black, very thick] (-3,3)--(-2,3)--(-2,5);
	\draw[black, very thick] (-1,2)--(0,2)--(0,6)--(-1,6);
	\draw[black, very thick] (-2,3)--(0,3);
	\draw[black, very thick] (-2,4)--(0,4);
	\draw[black, very thick] (-2,5)--(0,5);
	\draw[black, very thick] (0,6)--(1,6)--(1,7)--(0,7);
	\draw[black, very thick] (1,7)--(2,7);
	\draw[black, very thick] (0,2)--(1,2)--(1,1);
	\draw[black, very thick] (0,1)--(2,1);
	\draw (1,-1) node {\textbf{(b)}}; 
	\end{tikzpicture}
	\caption{(a) If, initially, the unit square centred at $(-n+1/2,-1/2)$ is covered by an horizontal domino (in grey), then all the dominoes adjacent to the northwest and southwest boundaries must also be horizontal (pink dominoes). The number of such tilings is $A_{n-1}$. (b) if the grey dominoes are fixed, then all the other dominoes adjacent to the northwest ans southwest boundaries must also be horizontal. The number of configurations is $A_{n-1}-A_{n-2}$. Indeed, with those fixed dominoes, we are left with an \textbf{AD} of order $n-1$ whose two leftmost squares are excluded or equivalently covered by a vertical domino. Hence the number of configuration is $A_{n-1}$ from which we must subtract the number of configurations (of an \textbf{AD} of order $n-1$) with two horizontal dominoes covering its leftmost part, which is $A_{n-2}$.}
	\label{fig:constraintsAD}
\end{figure}
The number of configurations satisfying this contraint is exactly $A_{n-1}$; hence, the probability $p$ to observe one of them becomes negligible as $n$ gets larger since it decreases exponentially with $n$:
\begin{equation}
p = \frac{A_{n-1}}{A_n} = 2^{-n} \xrightarrow[n\rightarrow +\infty ]{} 0.
\end{equation}
In other words, with probability $1$, the unit square centred at $(-n+1/2,-1/2)$ will be covered by a vertical domino. Because the \textbf{AD} is invariant under a quarter tour rotation, the unit square centred at $(n-1/2,-1/2)$ will also be covered, almost surely, by a vertical domino while the unit squares centred at $(-1/2,n-1/2)$ and $(-1/2,-n+1/2)$ will be covered by an horizontal domino. Let us push the reasoning a little farther to grasp the genesis of the arctic phenomenon, see Figure \ref{fig:constraintsAD} (b). The probability to have four horizontal dominoes with their left unit square centred at $(-n+3/2,-j+1/2)$, with $j=-2,-1,0,1$, also vanishes in the limit $n\rightarrow + \infty$, since the probability to observe a configuration satisfying these requirements is:
\begin{equation}
\frac{A_{n-1}-A_{n-2}}{A_n} \xrightarrow[n\rightarrow +\infty ]{} 0.
\end{equation}
Hence, with probability $1$, there must also be (at least) one vertical domino covering a unit square centred at $(-n+3/2,-j+1/2)$ for some $j\in \{-2,-1,0,1\}$. This heuristic suggests the emergence of brickwall patterns within each corner. Maybe unexpectedely, these frozen regions cover a non-negligible fraction of the \textbf{AD} (about $21.5\%$), even in the limit $n \rightarrow + \infty$. 
\section*{Non-intersecting lattice paths}
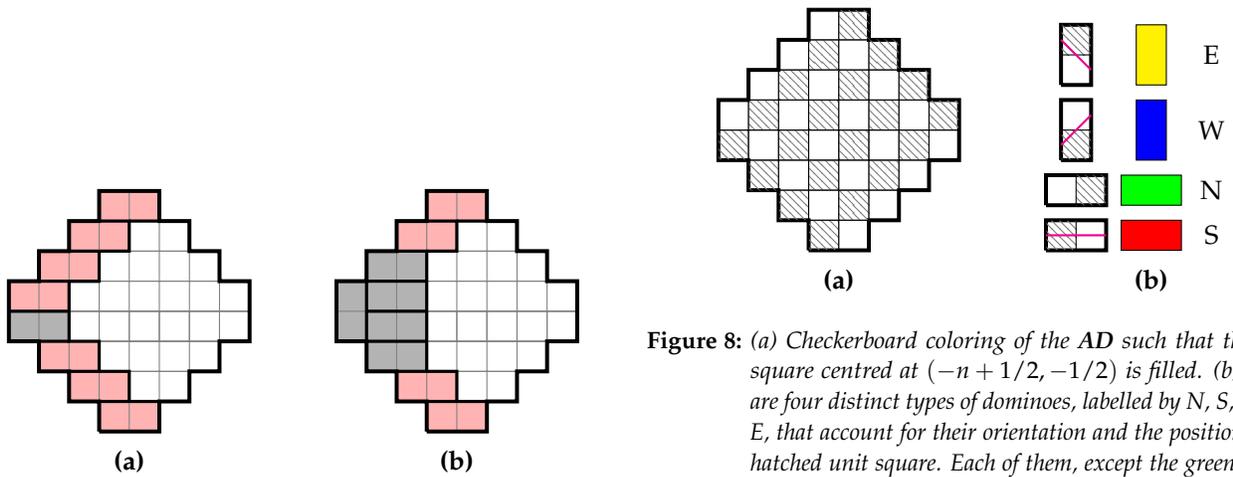
\begin{figure}
	\centering
	\begin{tikzpicture}[scale=0.4]
	\foreach \i in {0,1,2,3}
	{
		\draw[step=1cm,color=gray] (-\i,\i) grid (\i+2,\i+1);
		\draw[step=1cm,color=gray] (-3+\i,4+\i) grid (3-\i+2,4+\i+1);
	}
	\draw[step=1cm, draw=gray, fill=black!20!white] (-3,3) grid (3+2,3+1);
	\draw[color=black,line width=1.5pt] (0,0)--(2,0)--(2,1)--(3,1)--(3,2)--(4,2)--(4,3)--(5,3)--(5,5)--(4,5)--(4,6)--(3,6)--
	(3,7)--(2,7)--(2,8)--(0,8)--(0,7)--(-1,7)--(-1,6)--(-2,6)--(-2,5)--(-3,5)--(-3,3)--
	(-2,3)--(-2,2)--(-1,2)--(-1,1)--(0,1)--(0,0);
	\foreach \j in {0, ..., 4}
	\foreach \i in {0,1,2,3}
	{
		\draw[pattern=north west lines, pattern color=gray] (-\i+\j,\i+\j) rectangle (-\i+\j+1,\i+\j+1);
	}
{}	\draw (1,-1) node {\textbf{(a)}};
	\end{tikzpicture}\hspace{1cm}
	\begin{tikzpicture}[scale=0.4]
	\draw[color=black, line width=1.5] (0,0)--(2,0)--(2,1)--(0,1)--(0,0);
	\draw[pattern=north west lines, pattern color=gray] (0,0) rectangle (1,1);
	\draw[fill = red] (2.5,0) rectangle (4.5,1);
	\draw[black,magenta,thick] (0,0.5) -- (2,0.5);
	\node[color=black] (S) at (5.5, 1/2) {S};
	\foreach \i in {1.5}
	{
		\draw[color=black, line width=1.5] (0,0+\i)--(2,0+\i)--(2,1+\i)--(0,1+\i)--(0,0+\i);
		\draw[pattern=north west lines, pattern color=gray] (1,0+\i) rectangle (2,1+\i);
		\draw[fill = green] (2.5,0+\i) rectangle (4.5,1+\i);
		\node[color=black] (N) at (5.5, 1/2+\i) {N};
	}
	\foreach \i in {3}
	{
		\draw[color=black, line width=1.5] (0.5,\i)--(1.5,\i)--(1.5,2+\i)--(0.5,2+\i)--(0.5,\i);
		\draw[pattern=north west lines, pattern color=gray] (0.5,\i) rectangle (1.5,1+\i);
		\draw[fill = blue] (3,\i) rectangle (4,2+\i);
		\draw[black,magenta, thick] (0.5,\i+0.5) -- (1.5,\i+1.5);
		\node[color=black] (W) at (5.5, 1+\i) {W};
	}
	\foreach \i in {5.5}
	{
		\draw[color=black, line width=1.5] (0.5,\i)--(1.5,\i)--(1.5,2+\i)--(0.5,2+\i)--(0.5,\i);
		\draw[pattern=north west lines, pattern color=gray] (0.5,1+\i) rectangle (1.5,2+\i);
		\draw[fill = yellow] (3,\i) rectangle (4,2+\i);
		\draw[black,magenta, thick] (0.5,\i+1.5) -- (1.5,\i+0.5);
		\node[color=black] (E) at (5.5, 1+\i) {E};
	}
	\draw (3.5,-1) node {\textbf{(b)}};
	\end{tikzpicture}
	\caption{(a) Checkerboard coloring of the \textbf{AD} such that the unit square centred at $(-n+1/2,-1/2)$ is filled. (b) There are four distinct types of dominoes, labelled by N, S, W and E, that account for their orientation and the position of the hatched unit square. Each of them, except the green one, is associated with an elementary step (shown in magenta). }
	\label{fig:elementary_steps}
\end{figure}

In order to further characterise this arctic phenomenon, it is convenient to introduce an equivalent description in terms of non-intersecting lattice paths. The latter is obtained by first considering a checkerboard coloring of the domain, see Figure \ref{fig:elementary_steps} (a). 

\begin{figure*}
	\centering
	\begin{subfigure}[b]{0.42\textwidth}
		\includegraphics[width=\columnwidth]{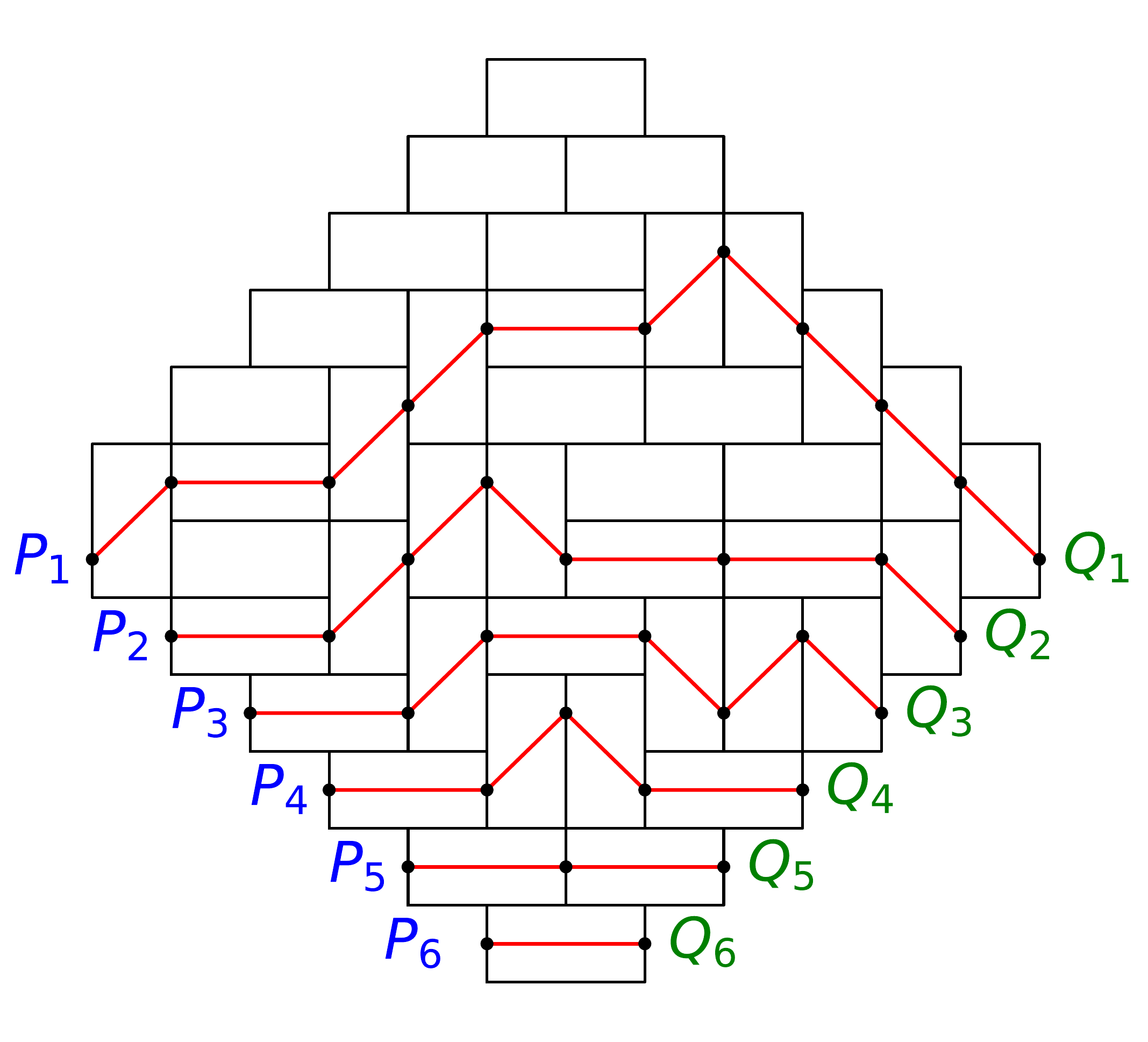}
		\caption{}
	\end{subfigure}\hspace{1cm}
	\begin{subfigure}[b]{0.38\textwidth}
		\includegraphics[width=\columnwidth]{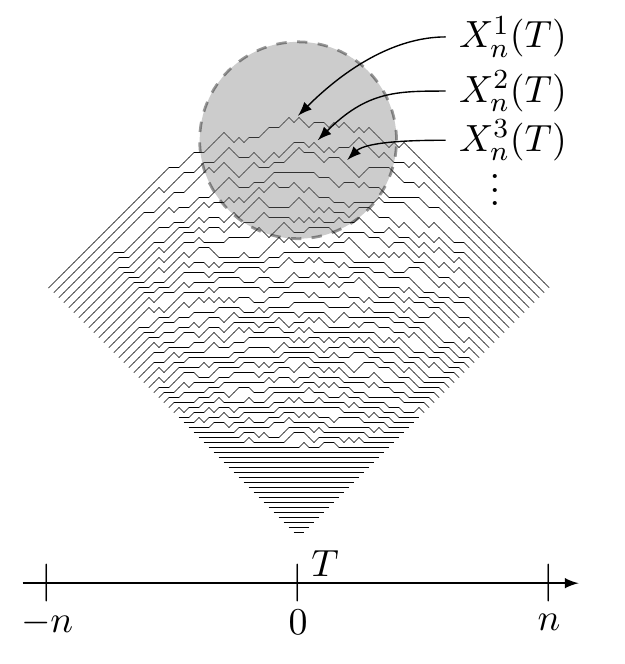}
		\caption{}
	\end{subfigure}\hspace{1cm}
	\caption{(a) A configuration of an \textbf{AD} of order $6$, along with its bijection in terms of non-intersecting lattice paths. (b) The same, for an \textbf{AD} of order $50$. The vertical position of the $k^{\text{th}}$ uppermost path is denoted by $X_n^k(T)$ for $-n+k-1 \leq T\leq n-k+1$. }
	\label{fig:bijection_AD_NILP}
\end{figure*}

This enables to distinguish four types of dominoes, labelled by a capital letter, according to the position of the hatched unit square, see Figure \ref{fig:elementary_steps} (b). Then, we draw on each domino, except for the N-dominoes which remain empty, a line segment as follows: the S-domino carries a horizontal $(2,0)$ line segment, the W-domino a diagonal $(1,1)$ line segment and the E-domino a diagonal $(1,-1)$ line segment. When the segments are actually drawn on the dominoes as prescribed, they form continuous paths which go across the diamond from the southwest boundary to the southeast boundary without intersecting. This procedure links, bijectively, each tiling of an \textbf{AD} of order $n$ to a set of $n$ non-intersecting paths (\cite{johansson2002NILP_AD}, section 2.1). For instance, Figure \ref{fig:bijection_AD_NILP} (a) shows a configuration of order $6$ along with its bijection in terms of non-intersecting lattice paths.

\begin{figure}
	\centering
	\includegraphics[width=\columnwidth]{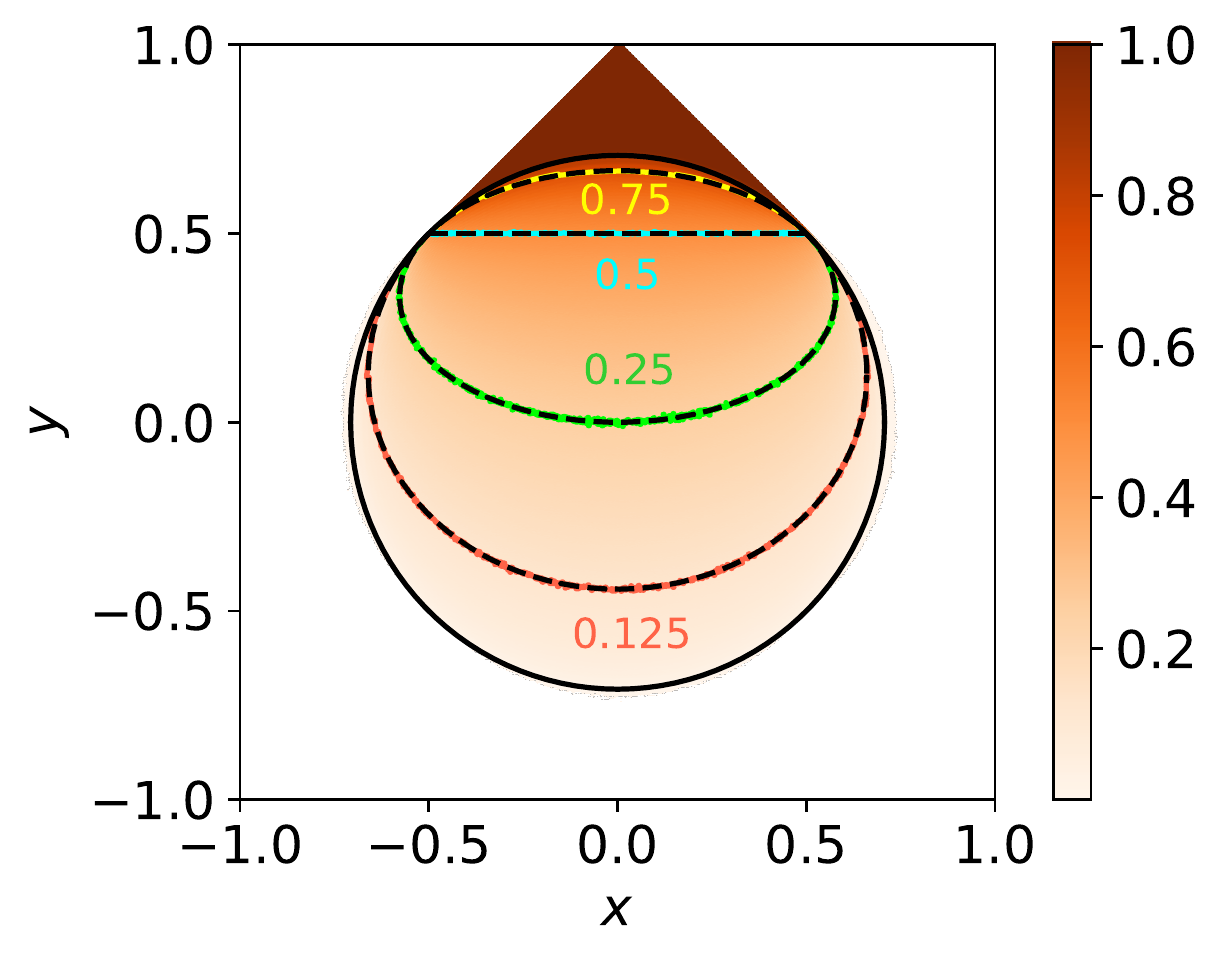}
	\caption{Density of N-dominoes, obtained from a sample of $100\,000$ configurations of order $n=500$. Lengths were divided by $n$. The arctic circle is shown in black. The predicted density level curves (dashed curves) give the set of points (i.e. ellipses) for which the density of N-dominoes is equal to $0.125, 0.25, 0.5, 0.75$. These curves agree with the numerical results shown in color.}
	\label{fig:density}
\end{figure}

As suggested by Figures \ref{fig:config_AD} and \ref{fig:bijection_AD_NILP} (b), the frozen north region is made up exclusively of N-dominoes and hence void of paths. In the disordered region, it was proved that the level curves for the density of N-dominoes (namely the set of points for which the density of N-dominoes is the same) are (incomplete) ellipses \cite{cohn2000localstat_ad}, as substantiated in Figure \ref{fig:density}: the closer to the north region and the larger the density of N-dominoes.

\section*{Tracy-Widom distribution and Gaussian Unitary ensemble}

\begin{figure*}
	\centering
	\begin{subfigure}[b]{\columnwidth}
		\centering
		\includegraphics[width=\columnwidth]{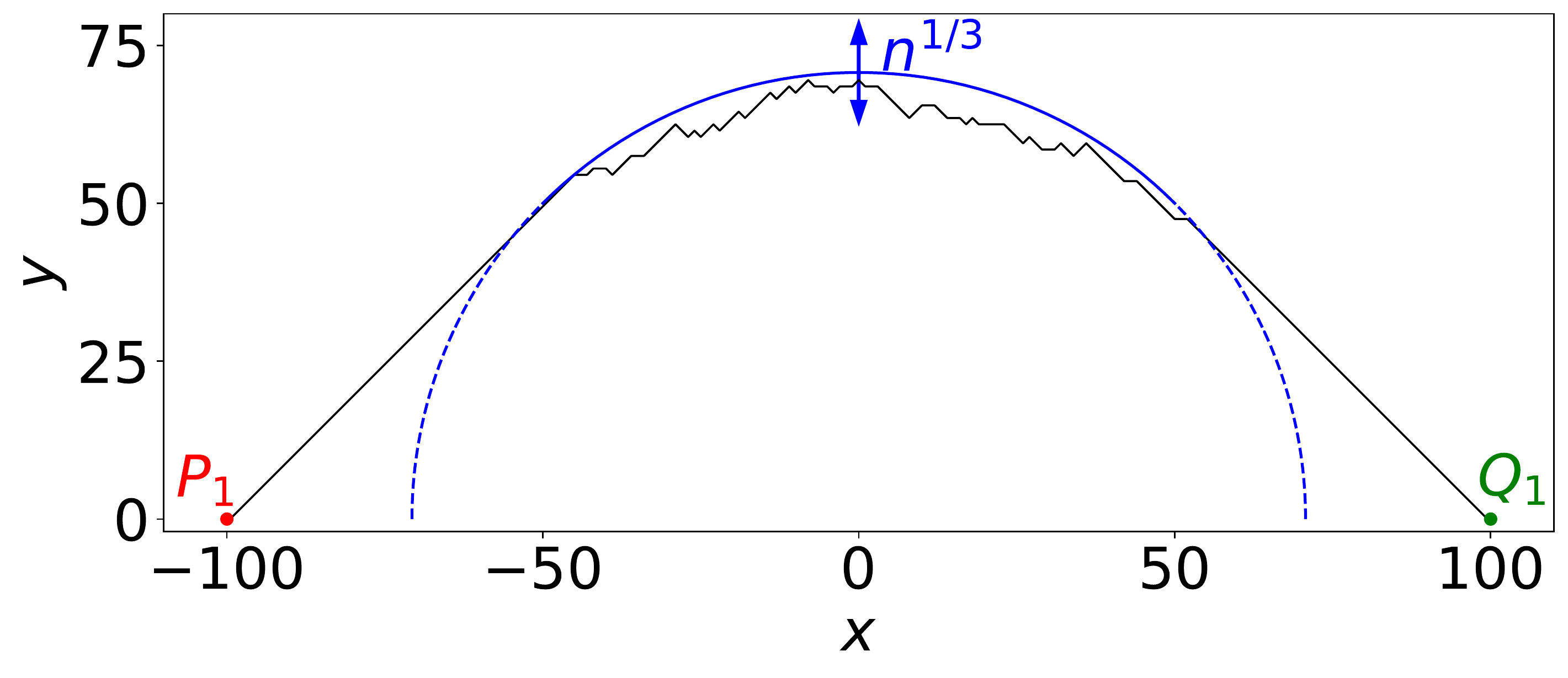}
		\caption{}
	\end{subfigure}
	\begin{subfigure}[b]{\columnwidth}
		\centering
		\includegraphics[width=\columnwidth]{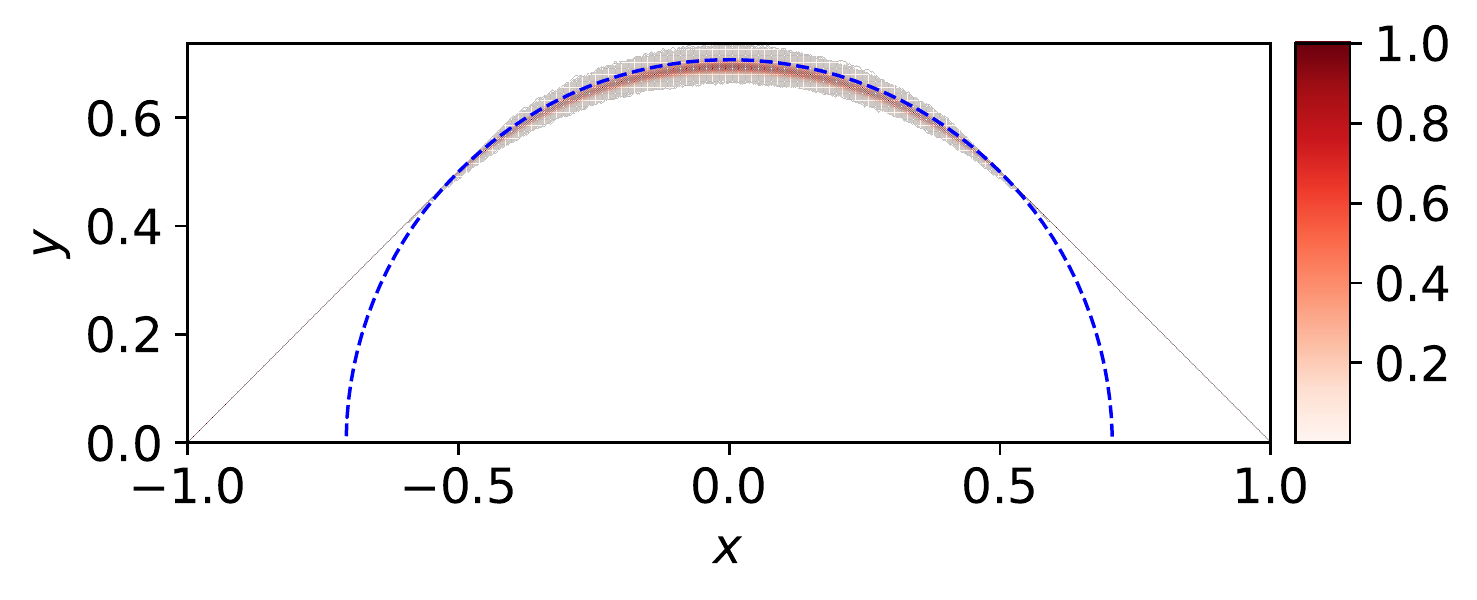}
		\caption{}
	\end{subfigure}
	\caption{(a) Uppermost path (black line) of an \textbf{AD} of order $n=100$, sampled with the shuffling algorithm. The uppermost path converges with probability $1$ to a quarter circle. The standard deviation between the uppermost path and the arctic circle, in a neighborhood of $X=0$, is proportional to $n^{1/3}$. (b) Probability density to observe the uppermost path in the rescaled domain (lengths divided by $n$), averaged over $100\,000$ indepedent configurations of order $500$.}
	\label{fig:arctic_circle}
\end{figure*}

Let $X_n^k(T)$ denote the vertical position of the $k^{\text{th}}$ topmost path at abscissa $T$ $(-n+k-1 \leq  T \leq n-k+1)$, for $k=1,\cdots,n$. The uppermost path $X_n^1(T)$ separates the central disordered region from the frozen north region. %
Configurations shown in Figure \ref{fig:arctic_circle} and \ref{fig:arctic_circle} (a) suggest the following convergence in probability:
\begin{equation}
\begin{split}
&\frac{X_n^1(T)}{n} \xrightarrow[n\rightarrow +\infty ]{\mathbb{P}} 1+\frac{T}{n} \quad \forall\, \frac{T}{n} \in [-1,-1/2[\\
&\frac{X_n^1(T)}{n} \xrightarrow[n\rightarrow +\infty ]{\mathbb{P}} \sqrt{\frac{1}{2}-{\Big(\frac{T}{n}\Big)}^2} \quad \forall\, \frac{T}{n} \in [-1/2,1/2[\\
&\frac{X_n^1(T)}{n} \xrightarrow[n\rightarrow +\infty ]{\mathbb{P}} 1-\frac{T}{n} \quad \forall\, \frac{T}{n} \in ]1/2,1]
\end{split}
\end{equation} 
In other words, as $n\rightarrow + \infty$ and after dividing all the lengths by $n$, the uppermost path converges almost surely to a straight line segment from $(-1,0)$ to $(-1/2,1/2)$, a quarter circle from $(-1/2,1/2)$ to $(1/2,1/2)$ and again a straight line segment from $(1/2,1/2)$ to $(1,0)$.

In particular, for $T=0$, we have:
\begin{equation}
\frac{X_n^1(0)}{n} \xrightarrow[n\rightarrow +\infty ]{\mathbb{P}} \frac{1}{\sqrt{2}}
\end{equation}
This is however not the whole story. If we generate a vast collection of tilings and report the vertical position $X_n^1(0)$ for each of them, then we obtain a set of points whose distance from $\frac{n}{\sqrt{2}}$ is proportional, on average, to $n^{1/3}$, see Figure \ref{fig:arctic_circle}. In other words:
\begin{equation}
\rev{\mathbb{E}\Big(X_n^1(0)-\frac{n}{\sqrt{2}}\Big) \sim n^{1/3}~~ \text{as } n \rightarrow + \infty,}
\end{equation}   
where the expected value $\mathbb{E}$ is taken over the whole set of tilings of \textbf{AD} of order $n$. The next step is to determine which probability distribution $\xi$ (as far as it exists) governs the fluctuations of the properly rescaled vertical position $X_n^1(0)$:
\begin{equation}
\lim_{n\rightarrow + \infty} \mathbb{P} \Big(
\frac{X_n^1(0)-\frac{n}{\sqrt{2}}
}{
n^{1/3}}
\leq s
\Big) = \xi(s),
\label{eq:Johansson_0}
\end{equation}
Let us decompose $X_n^1(0)$ as follows:
\begin{equation}
\begin{split}
&X_n^1(0) = X_n^n(0)+\sum_{k=1}^{n-1} Y_k,\\
&Y_k = X_n^{k}(0)-X_n^{k+1}(0).
\end{split}
\end{equation}
If the increments $Y_k$ were independent and identically distributed random variables, then $\xi$ would be the normal distribution \rev{(provided a scaling $n^{1/2}$ for the fluctuations is used)}, by virtue of the central limit theorem. Here, however, the increments are not independent and identically distributed. Indeed, \JFadd{based on heuristic arguments (see section $3.2$ in \cite{debin2021exploration})}, we have with probability $1$ that $X_n^{k}(0)-X_n^{k+1}(0)=O(n^{1/3})$ for $k = O(1)$ while $X_n^{k}(0)-X_n^{k+1}(0)=1$ almost surely in the south frozen region, see Figure \ref{fig:bijection_AD_NILP} (b). Hence, we should not expect $\xi$ to be a Gaussian distribution. Actually, the distribution $\xi$ is explicitly known and is given by \cite{johansson2005_airy}:
\begin{equation}
\xi(s) = F_2^1(2^{5/6}s),
\end{equation}
with $F_2^1$ the Tracy-Widom \rev{cumulative} distribution\footnote{The Tracy-Widom distribution $F_2^1$ can be expressed either as a Fredholm determinant or in terms of the solution of Painlevé II equations \cite{bornemann2009numerical_fred_review}.}, see Figure \ref{fig:conjecture}.
The Tracy-Widom was first introduced in random matrix theory to describe the fluctuations of the largest eigenvalue of matrices belonging to the Gaussian Unitary Ensemble \cite{tracy1994tw_distr}. This ensemble is the set of $n\times n$ hermitian matrices $H$ whose entries are Gaussian variables distributed as follows:
\begin{equation}
\begin{split}
&H_{kk} \sim N(0,1/2),\\
&H_{kl} \sim N(0,1/4) + i N(0,1/4) \quad (1\leq k<l \leq n).
\end{split}
\end{equation}
The joint probability density function $P(\lambda_1,\lambda_2,\cdots,\lambda_n)$ of the ordered eigenvalues $\lambda_1 \geq \lambda_2 \geq \cdots \geq \lambda_n$ is given by:
\begin{equation}
\begin{split}
\mathbb{P}(\lambda_1,\lambda_2,\cdots,\lambda_n)
&=\frac{1}{Z_n} 
\prod_{1\leq i < j \leq n} {\vert \lambda_i - \lambda_j \vert}^2 
e^{-\sum_{i=1}^n \lambda_i^2}\\
&=\frac{1}{Z_n} 
e^{\sum_{i\neq j} \log \vert \lambda_i - \lambda_j \vert -\sum_{i=1}^n \lambda_i^2},
\end{split}
\label{Coulomb_gas}
\end{equation}
for some normalisation constant $Z_n$. Two competing terms are at play in the above expression: the term $\vert \lambda_i - \lambda_j \vert$ \rev{indicates} the repulsion of eigenvalues while the term $e^{-\sum_{i=1}^n \lambda_i^2}$ favours the attraction \rev{of the eigenvalues towards the origin.} 
The rescaled largest eigenvalues:
\begin{equation}
\Lambda_n^i := \frac{\lambda_i-2\sqrt{n}}{2^{-1/2}n^{-1/6}},
\end{equation} 
are random variables whose \rev{cumulative distribution functions}, in the limit $n\rightarrow + \infty$, are denoted $F_2^i$ and shown in Figure \ref{fig:conjecture} for $i=1,\cdots,6$. The Tracy-Widom distribution appears in many contexts: without being exhaustive, it is found in combinatorics \cite{baik1999longest_increasing_seq} and in some growth processes, such as the polynuclear growth model \cite{prahofer2000universal}. It was also found experimentally by applying an alternative current voltage to a thin layer of nematic liquid crystal. For sufficiently large values of the voltage, two distinct turbulent phases coexist. The interface between the two phases grows with time and its fluctuations were shown to obey the Tracy-Widom cumulative distribution function \cite{takeuchi2011growing}. The common point of all these models is the unusual scale $n^{1/3}$ of the fluctuations where $n$ provides some measure of the size of the system (such as the order of the \textbf{AD}). Indeed writing $\Lambda^i_n$ as
\begin{equation}
\Lambda_n^i = \frac{\sqrt{n} \lambda_i - 2n}{2^{-1/2}n^{1/3}}
\end{equation}
makes $X^1_n(0)$ and $\sqrt{n} \lambda_i$ comparable: their average and deviations have the same scaling law with $n$.

\section*{Airy line ensemble}

So far, we have seen that the largest eigenvalue $\lambda_1$ of a random hermitian matrix and the vertical position $X_n^1(0)$ of the uppermost path of an \textbf{AD} are both random variables governed by the same probability distribution in the limit of large $n$. A stronger result was actually proved in \cite{johansson2005_airy} where it is shown that the uppermost path $X_n^1(T)$ plus a quadratic term converges (in the sense of finite-dimensional distributions and after a proper rescaling) to the Airy process $A_2^1(t)$, with $t=2^{1/6}n^{-2/3}T$. The Airy process $A_2^1(t)$ is a stationary and translation-invariant process whose \rev{cumulative distribution function}, at fixed $t$, is the Tracy-Widom distribution $F_2^1$. Hence, the result \JFadd{proved in \cite{johansson2005_airy}} implies:
\begin{equation}
 \begin{split}
 \lim_{n \to \infty} &\mathbb{P}   \left(\frac{X_n^1(2^{-1/6}n^{2/3}t)-\frac{n}{\sqrt{2}}}{2^{-5/6}n^{1/3}}+t^2\leq s \right)  \\
 &= \mathbb{P}(\mathcal{A}_2^1(t)\leq s)= F_2^1(s).
 \label{eq_AD_TW_1}
 \end{split}
\end{equation}
In particular, for $t=0$, we recover eq. (\ref{eq:Johansson_0}).

\begin{figure}
	\centering
	\includegraphics[scale = 0.35]{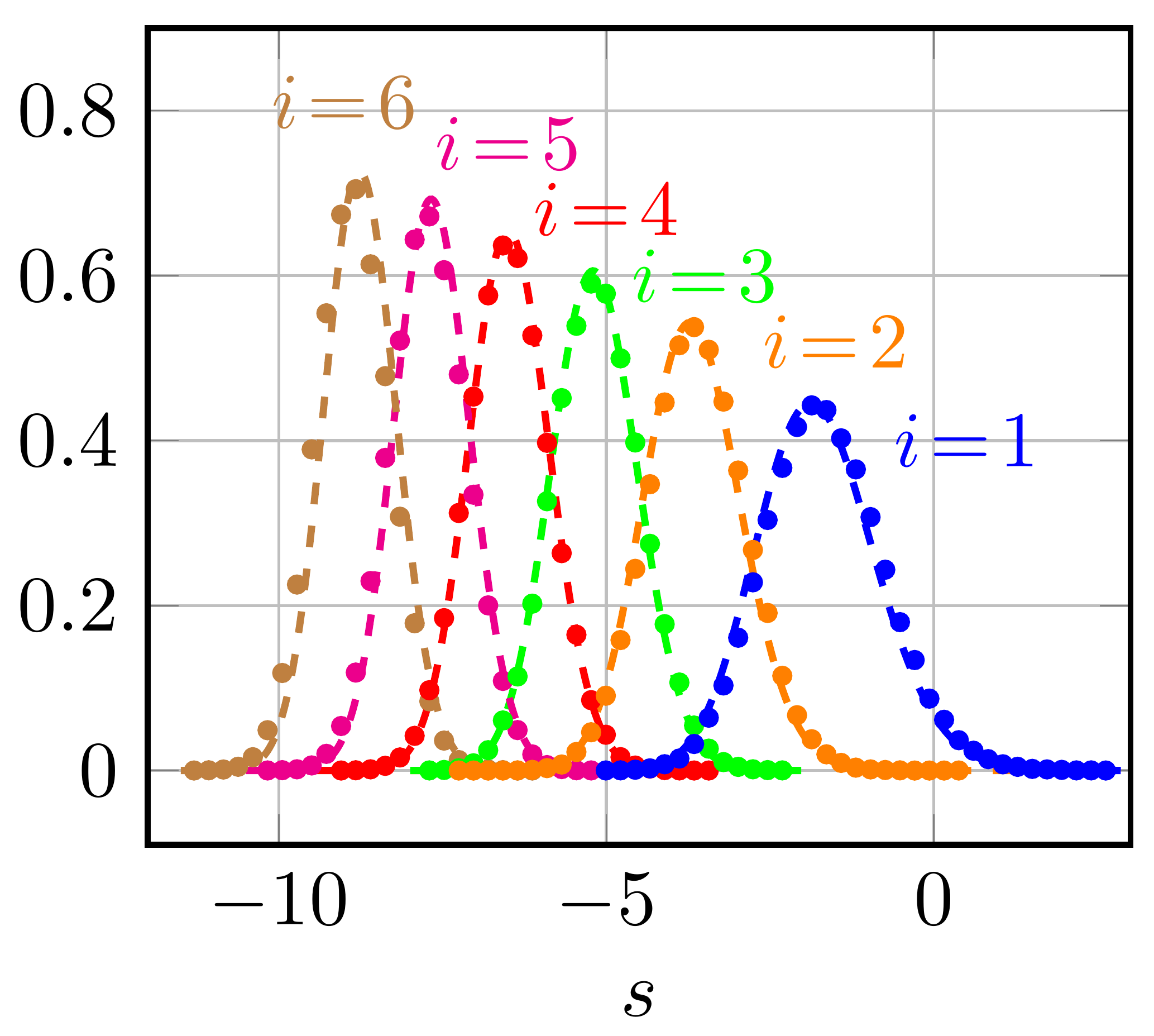}
	\caption{The dashed curves give the probability density functions \rev{$\frac{d}{ds}F_2^i(s)$} associated with the $i^{\text{th}}$ largest eigenvalue of random hermitian matrices sampled from the Gaussian Unitary Ensemble. The points give %
		$
		\frac{\mbox{d}}{\mbox{d}s}\mathbb{P}   \left(\frac{X_n^i(2^{-1/6}n^{2/3}t)-\frac{n}{\sqrt{2}}}{2^{-5/6}n^{1/3}}+t^2\leq s \right)
		$, for the $6$ uppermost paths ($i=1,\cdots,6$) of an \textbf{AD} or order $n=500$ and $T=-1/2$ (corresponding to $t\approx -0.0089$). The data were obtained from a sample of $100\,000$ configurations.}
	\label{fig:conjecture}
\end{figure}

Quite naturally, the following question arises: does this correspondence also holds for the other eigenvalues and paths, i.e. do $X_n^k(T)$ and $\sqrt{n} \lambda_k$ have the same limiting probability distributions, for $k=2,3,\cdots$ ($k=O(1)$) and fixed $T$ ? Although to the best of our knowledge no analytic proof has been provided for the limiting probability distributions of $X_n^k(T)$ ($k=2,3,\cdots$), heuristic arguments supported by numerical simulations clearly point to a positive answer. Roughly, in a neighborhood of $t=0$, the uppermost paths of an \textbf{AD} look like Brownian motions constrained not to cross, see Figure \ref{fig:bijection_AD_NILP} (b). This is reminiscent of the \textit{time-dependent Coulomb gas model} considered in \cite{dyson1962brownian} and of Brownian bridges constrained not to intersect and with identical starting and ending points \cite{corwin2014brownian}. In the latter case, it was shown that the set of curves plus a quadratic term converges, under a proper rescaling, to the Airy line ensemble. Based on the similarities between those models and the description in terms of non-intersecting lattice paths of an \textbf{AD}, we conjecture that the uppermost paths of an \textbf{AD} should also converge to the Airy line ensemble, i.e.:
\begin{equation}
\begin{split}
\lim_{n \to \infty} &\mathbb{P}   \left(\frac{X_n^i(2^{-1/6}n^{2/3}t)-\frac{n}{\sqrt{2}}}{2^{-5/6}n^{1/3}}+t^2\leq s \right)  \\
&= \mathbb{P}(\mathcal{A}_2^i(t)\leq s)= F_2^i(s),
\label{eq_AD_TW_i}
\end{split}
\end{equation}
for $i=O(1)$. This conjecture is well supported by Figure \ref{fig:conjecture}, where we compare, for $T=-1/2$, the probability density functions of the first few uppermost paths under the rescaling given in eq. \ref{eq_AD_TW_i} with the \rev{probability density functions} $\frac{d}{ds}F_2^i(s)$ ($i=1,\cdots,6$).

\section*{Conclusion}
In this work, we have considered domino tilings of planar regions. We have seen that the boundary of the domain can drastically change the properties of the model and lead to an arctic phenomenon, as it is the case for the celebrated Aztec diamond. In the limit of large domain and under a proper rescaling, it is known that the boundary between the frozen region and the central disordered one converges to the Airy process whose $1$-point distribution function is the Tracy-Widom distribution for the largest eigenvalue of random hermitian matrices. Based on heuristic arguments and numerical simulations, we conjecture that the boundary should converge to the Airy line ensemble, which extends the Airy process. 

\section*{Acknowledgments}
\JFadd{The authors thank Timoteo Carletti for the careful reading of the manuscript.} Part of the results were obtained using the computational resources provided by the ``Consortium des Equipements de Calcul Intensif'' (CECI), funded by the Fonds de la Recherche Scientifique de Belgique (F.R.S.-FNRS) under Grant No. 2.5020.11 and by the Walloon Region. JF DK is supported by a FNRS Aspirant Fellowship under the Grant FC38477. PR is Senior Research Associate of \rev{the} FRS-FNRS (Belgian Fund for Scientific Research).    


\bibliographystyle{plain}
\bibliography{biblio_v3}


\end{document}